\documentclass[twocolumn,amsmath,prb]{revtex4}

\usepackage{graphicx}

\begin{document}

\def\bq{{\bf q}}
\def\br{{\bf r}}
\def\bS{{\bf S}}
\def\d{\delta}
\def\f{\frac}
\def\l{\lambda}
\def\p{\partial}
\def\s{\sigma}
\def\v{\varepsilon}
\def\tI{{\tilde{I}}}
\def\tJ{{\tilde{J}}}
\def\tS{{\tilde{S}}}
\def\cN{{\cal{N}}}

\title{
  Instability of planar vortices in two-dimensional easy-plane Heisenberg model
  with distance-dependent interactions 
}
\author{Myoung Won Cho}
\email{mwcho@postech.edu}
\author{Seunghwan Kim}
\email{swan@postech.edu}
\affiliation{
  Asia Pacific Center for Theoretical Physics $\&$ NCSL,
  Department of Physics, Pohang University of Science and Technology,
  Kyungpook, Pohang, 790-784, South Korea}
\date{\today}

\begin{abstract}
It is known that magnetic vortices in two dimensional Heisenberg models with
easy-plane anisotropy exhibit an instability depending on the anisotropy
strength. 
In this paper, we study the statistic behavior of the two-dimensional
easy-plane Heisenberg models with distance-dependent interactions, $J_{xy}(r)$
and $J_z(r)$ for in-plane and out-of-plane components.
We develop analytical and numerical methods for accurate determination of
critical anisotropy, above which out-of-plane vortices are stable.
In particular, we explore the vortex formation of the Gaussian-type interaction
model and determine the critical anisotropy accurately for square, hexagonal
and triangular lattices.

\end{abstract}

\maketitle

\renewcommand{\theequation}{\arabic{section}.\arabic{equation}}

%
%
\section{Introduction}
\setcounter{equation}{0}

The magnetism in two-dimension has been the subject of continued interest in
the last few decades, which is aroused further due to recent additions of new
materials such as Cu-based high-temperature superconductors and advances in
both numerical and experimental capabilities.
The classical two-dimensional anisotropic Heisenberg model (CTDAHM) with the
spin Hamiltonian,
\begin{eqnarray} \label{eq:CTDAHM}
  H=-K\sum_{\langle i,j\rangle}\left(S_i^xS_j^x+S_i^yS_j^y
    +\l S_i^zS_j^z\right)
\end{eqnarray}
for a coupling constant $K$, has been one of the most studied models for
magnetic systems.
For $\l=0$, it becomes the so-called XY model, which is well understood
following the work of Berezinskii~\cite{Berezinskii1971}
and Kosterlitz and Thouless~\cite{Kosterlitz1973}.
It has a phase transition at $T_{KT}$, named the Kosterlitz-Thouless phase
transition, which can be characterized by a vortex-antivortex
unbinding~\cite{Kosterlitz1974}.
For $\l<1$, it belongs to the universality class of the XY model, and it has
been known that there are two different types of nonlinear solutions, called
``in-plane'' and ``out-of-plane'' vortices~\cite{Wysin1988}.
There have been several attempts with numerical simulations and analytic
calculations to obtain the critical anisotropy $\l_c$, above which the
out-of-plane vortices are stabilized.
At first, Gouv{\^e}a {\em et al.}~\cite{Gouvea1989} investigated the
development of the out-of-plane vortex by using a combined Monte
Carlo-molecular-dynamics technique.
They found that initial out-of-plane vortex relaxes to a planar one below
the critical anisotropy $\l\approx 0.72$, $0.86$, $0.62$ for square, hexagonal
and triangular lattices respectively.
Costa and Costa~\cite{Costa1996} obtained a more precise measurement
$\l_c\approx 0.709$ for a square lattice by fitting the out-of-plane squared
component $(S_z)^2$ at the central peak as a function of $\l$.
Wysin made an analysis of the core region of a vortex on a discrete lattice
and obtained an estimate of $\l_c\approx 0.7044$ for the square
lattice~\cite{Wysin1994}, which is improved further to a very accurate value
of $\l_c\approx 0.703409$ by iteratively setting each spin's xy components to
point along the direction of the effective field due to its neighbors.

Recently, we showed that the pattern formation in primary visual cortex is
also related with the vortex dynamics in magnetism~\cite{Cho2003A}.
It is observed {\em in vivo} that the development of strong scalar components
(ocular dominance column) near the singularity centers at orientational
components (orientation column).
There are at least three different types in observed ocular dominance patterns
{\em in vivo}, which correspond to the Ising type, stable out-of-plane vortex,
and stable in-plane vortex behaviors~\cite{Cho2003B}.
The determination of the critical anisotropy is important in describng the
region of different behaviors in vortex patterns.

In this paper, we consider the two-dimensional easy-plane (XY) symmetry
Heisenberg model with distance-dependent interactions :
\begin{eqnarray} \label{eq:H}
  H=-\sum_{i,j}\left\{J_{xy}(r_{ij})(S_i^xS_j^x+S_i^yS_j^y)
    +J_z(r_{ij})S_i^zS_j^z\right\} \ \ 
\end{eqnarray}
for the classical spin variables $\bS_i=(S_i^x,S_i^y,S_i^z)$ and the distance
dependent exchange energy $J_\mu(r)=\v_\mu I_\mu(r)/2$ ($\mu=xy$ or $z$).
The dimensionless neighborhood interaction function $I_\mu(r)$ can have a
various shape but is expected to be a smooth function and approach zero as
$r\rightarrow\infty$.
Comparing Eq.(\ref{eq:H}) with Eq.(\ref{eq:CTDAHM}), we may expect that the
anisotropy in the strength of interactions between in-plane and out-of-plane
components is $\l\sim J_z/J_{xy}$ and obviously $\l=\v_z/\v_{xy}$ if
$I_{xy}(r)=I_z(r)$ for all $r$.
But a more clear definition of the anisotropy parameter $\l$ will be needed 
for the case of arbitrary interaction functions with $I_{xy}(r)\neq I_z(r)$.
Takeno and Homma showed that the anisotropic Heisenberg model with
distance-dependent interactions has two different minima in $H$ depending on
the ground-state energy of in-plane and out-of-plane
components~\cite{Takeno1980}.
We determine the anisotropy $\l$ by the ratio of the ground-state energy, which
is the boundary of the crossover behavior between the XY and Ising models.
The critical anisotropy now may depend on the shape and parameter values of
$J_\mu(r)$.

In Sec.~\ref{sec:effective}, we build an effective Hamiltonian in a continuum
approximation for the arbitrary exchange energy function $J(r)$.
The continuum theory helps to predict or understand the major behavior of spin 
configurations and vortex formation.
But the exact estimation of the critical anisotropy cannot be obtained by a
continuum approximation because of the singularity near the vortex
core~\cite{Wysin1994}.
Sometimes two distinct spin models can yield the same Hamiltonian in a
continuum approximation.
We will show that different values of the critical anisotropy $\l_c$ is
possible in spite of the common approximated Hamiltonian
(Sec.~\ref{sec:gaussian}).
In Sec.~\ref{sec:analytic}, we generalize and extend Wysin's
method~\cite{Wysin1994,Wysin1998}, which is the analytic method for obtaining
the exact critical anisotropy $\l_c$ in the distant neighbor interaction models.
We clarify our method by calculating critical anisotropies of the CTDAHM with
$I(r)=\delta(r-1)$ for different lattice types and comparing them to values.
In Sec.~\ref{sec:simulations}, we exhibit a simulational method for the
determination of the critical value of anisotropy above which out-of-plane
components develop.
We find that the out-of-plane components start to develop for $\l\ge\l_c$ where
$\l_c$ are $0.7035$, $0.8330$ and $0.6129$ in four-digit accuracy for square,
hexagonal and triangular lattices, respectively.
These values from simulations are the closest values to the analytic
calculations~\cite{Wysin1998}.
In Sec.~\ref{sec:gaussian}, we investigate the critical anisotropy in the
two-dimensional easy-plane Heisenberg model with Gaussian type interactions of
$I_\mu(r)=\exp(-r^2/2\s_\mu^2)$ and the anisotropy
$\l=(\v_z/\v_{xy})(\s_z^2/\s_{xy}^2)$.
We obtain the critical anisotropy $\l_c$ for $\l=\v_z/\v_{xy}$
($\s_{xy}=\s_{z}$) and $\l=\s_z^2/\s_{xy}^2$ ($\v_{xy}=\v_z$) depending on the
interaction ranges $\s_\mu$ for different lattice types and find a general
behavior of $\l_c$.

%
%
\section{Effective Hamiltonian approach} \label{sec:effective}
\setcounter{equation}{0}

The classical spin vector, $S_i=(S_i^x,S_i^y,S_i^z)$, can be specified by two
angles of rotation $\phi_i$ and $\theta_i$,
\begin{eqnarray} \label{eq:spin}
\bS_i=S(\sin\theta_i\cos\phi_i, \sin\theta_i\sin\phi_i, \cos\theta_i).
\end{eqnarray}
In the momentum space representation, we write
\begin{eqnarray} \label{eq:Sq}
  \tS_\bq=\f{1}{\sqrt{N}}\sum_i S_ie^{-i\bq\cdot\br_i},
\end{eqnarray}
where the vectors $\bq$ are restricted to the first Brillouin zone of the
simple cubic $d$-dimensional lattice.
Substituting Eqs.(\ref{eq:spin}) and (\ref{eq:Sq}) into the Hamiltonian in
Eq.(\ref{eq:H}), we get
\begin{eqnarray} \label{eq:tH}
H&=&-\sum_\bq\left\{\tJ_{xy}(\bq)(\tS_{q_x}\tS_{-q_x}
  +\tS_{q_y}\tS_{-q_y})\right. \nonumber \\
  && \hspace{2cm} \left.+\tJ_z(\bq)\tS_{q_z}\tS_{-q_z}\right\}
\end{eqnarray}
where
\begin{eqnarray} \label{eq:tJ}
  \tJ_\mu(\bq)&=&\sum_\br J_\mu(\br)e^{-i\bq\cdot\br}
\end{eqnarray}
for $\mu=xy$ or $z$.
If $\tJ_\mu(q)$ has a maximum point at $q=0$, the approximated Hamiltonian is
obtained by the inverse fourier transform of the expansion in Eq.(\ref{eq:tH}) :
\begin{eqnarray} \label{eq:approx}
H&\simeq&-NJ_sS^2+\f{J_sS^2}{4}\left.\int d^2r\right\{\alpha(1-m^2)(\nabla\phi)^2 \nonumber \\
  && \ \ \left.+\alpha[1-\d'(1-m^2)]\f{(\nabla m)^2}{(1-m^2)}+4\d m^2 \right\}
\end{eqnarray}
where $J_s=\tJ_{xy}(0)$, $J_p=-\tJ_{xy}''(0)/a^2$, $\d=1-\tJ_z(0)/\tJ_{xy}(0)$,
$\d'=1-\tJ_z''(0)/\tJ_{xy}''(0)$, $\alpha=2J_p/J_s$ and $m(\br)=S_z(\br)/S$
for the nearest-neighbor distance $a$.
The exchange energy $J_\mu(r)$ need not be fully positive, however, in
ferromagnetic behavior systems.
Comparing with the approximated Hamiltonian of the CTDAHM in continuum
theory~\cite{Gouvea1989}, we expect a ferromagnetic behavior if $J_s$ and $J_p$
are positive.
The anisotropy $\l$ is described as
\begin{eqnarray} \label{eq:anisotropy}
  \l=1-\d=J_z(0)/J_{xy}(0),
\end{eqnarray}
which is the ratio of the ground-state energy for in-plane and out-of-plane
components.
Eq.(\ref{eq:approx}) has two different solutions depending on the anisotropy
in $\l$.
The total ground-state energy in Eq.(\ref{eq:approx}), $\langle H\rangle_0$, is
given by $-N\tJ_{xy}(0)S^2$ (or $-N\tJ_z(0)S^2$) and $m^2(\br)=0$ (or
$m^2(\br)=1$) for all $\br$ when $\l<1$ (or $\l>1$), which is the
XY (or Ising)-like solution as shown by Takeno {\em et al.}~\cite{Takeno1980}
The anisotropy satisfies $\l=\v_z/\v_{xy}$ if $I_{xy}(r)=I_z(r)$ for all $r$.
However $I_{xy}(r)\neq I_z(r)$, the anisotropy can be expressed as
$\l\propto(\s_z/\s_{xy})^d$ for the interaction range $\s_\mu$ in the
$d$-dimensional lattice because the exchange energy is proportional to the
number of interaction pairs in ferromagnetic systems.

%
%
\section{Analytic calculations of critical anisotropy} \label{sec:analytic}
\setcounter{equation}{0}

The Hamiltonian in Eq.(\ref{eq:H}), can be rewritten as
\begin{eqnarray}
  H&=&-\sum_{i,j}\{J_{xy}(\br_{ij})\sqrt{1-m_i^2}\sqrt{1-m_j^2}\cos(\phi_i-\phi_j)
   \nonumber \\
  &&\ \ \ \ \ \ \ \ \ \ \ +J_z(\br_{ij})m_im_j \},
\end{eqnarray}
where $\phi_i=\tan^{-1}(S_i^y/S_i^x)$ and $m_i=S_i^z/S$.
For arbitrary $|m_i|\ll 1$, we get
\begin{eqnarray} \label{eq:Fij}
  \f{\p H}{\p m_i}&\simeq&2\sum_j\{J_{xy}(\br_{ij})m_i\cos(\phi_i-\phi_j) 
    \nonumber \\
    && \hspace{1.5cm} -J_z(\br_{ij})m_j\} = 0.
\end{eqnarray}
This always has an in-plane vortex solution, $m_i=0$ for all sites $i$.
For the out-of-plane solutions, which have non-zero $m_i$ components, the
determinant of the matrix $W$ should vanish, where
\begin{eqnarray} \label{eq:W_ab}
  W_{\alpha\beta}&=&\d_{\alpha\beta}\sum_{i\in M_\alpha}\sum_j
     J_{xy}(\br_{ij})\cos(\phi_i-\phi_j) \nonumber \\
    &&-\sum_{i\in M_\alpha}\sum_{j\in M_\beta}J_z(\br_{ij}).
\end{eqnarray}
For an arbitrary exchange energy function $J_\mu(r)$, $\mu=xy$ or $z$, the
critical value of the anisotropy can be determined by a root finding method for
$\det|W(\l)|=0$.
When $I_{xy}(r)=I_z(r)$ for all $r$, so $\l=\v_z/\v_{xy}$, the boundary for
the existence of the out-of-plane solutions can be obtained from the eigenvalue
problem,
\begin{eqnarray}
A_\alpha m_\alpha-\l w_{\alpha\beta}m_\beta = 0
\end{eqnarray}
or
\begin{eqnarray} \label{eq:eigenproblem}
(w_{\alpha\beta}/A_\alpha)m_\beta = (1/\l)m_\alpha
\end{eqnarray}
for
\begin{eqnarray}
A_\alpha=\f{1}{|M_\alpha|}\sum_{i\in M_\alpha}\sum_jI(\br_{ij})
  \cos(\phi_i-\phi_j)
\end{eqnarray}
and
\begin{eqnarray}
w_{\alpha\beta}=\f{1}{|M_\alpha|}\sum_{i\in M_\alpha}\sum_{j\in M_\beta}
  I(\br_{ij})
\end{eqnarray}
where $\alpha$ or $\beta$ is a block index rather than a site index when there
exists a symmetry in the lattice geometry and
$M_\alpha=\{\ i\ |\ r_i=r_\alpha\}$ are sets ordered by the distance from the
vortex core ($1\le\alpha\le L$).
The maximum eigenvalue, which is equal to $1/\l_c$, is determined by using only
the nearest components to the core center, $m_1,m_2,\ldots m_\cN$
($1\le\cN\le L$, where $\cN$ is the number of the nearest components).

\begin{figure}
\begin{minipage}[b]{6cm}
  \includegraphics[width=6cm]{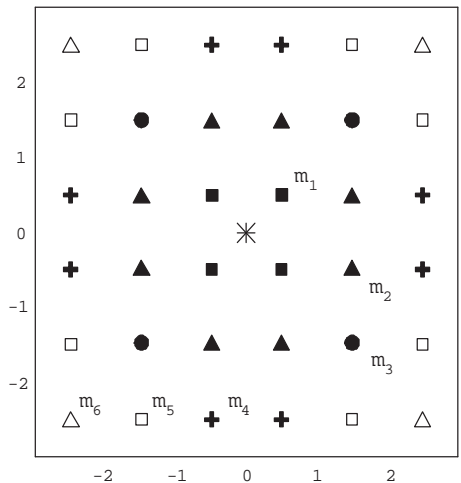}
\end{minipage}
\caption{ \label{fig:square}
  The vortex center is located at $(0, 0)$ and the sites are presented by
  different symbols according to the distance from the vortex center.
  (solid squares for $r=1/\sqrt{2}$, solid triangles for $r=\sqrt{10}/2$, etc.)
}
\end{figure}

Note that we assume that the center of the vortex is located at the origin of a
coordinate system.
The in-plane angles are chosen to be the usual solution of the in-plane vortex
in the continuum theory, $\phi_i=\tan^{-1}(y_i/x_i)$, or
\begin{eqnarray} \label{eq:phi_ij}
  \cos(\phi_i-\phi_j)=\hat{\br}_i\cdot\hat{\br}_j.
\end{eqnarray}
For the CTDAHM in Eq.(\ref{eq:CTDAHM}), the interaction function is given by
$I(\br)=\d(r-1)$ and
\begin{eqnarray}
  A_1&=&\f{4}{\sqrt{5}} \nonumber \\
  A_2&=&\f{4}{\sqrt{5}}(1+\f{1}{\sqrt{5}}+\f{2}{\sqrt{13}}) \\
  w_{11}&=&w_{12}\ =\ 2 \nonumber \\
  w_{21}&=&w_{22}\ =\ 1, \nonumber 
\end{eqnarray}
where $M_1=\{\pm(1/2,1/2),\pm(1/2,-1/2)\}$ and
$M_2=\{\pm(3/2,1/2)$,$\pm(3/2,-1/2)$,$\pm(1/2,3/2)$,$\pm(1/2,-3/2)\}$ for the
square lattice in Fig \ref{fig:square}.
In this case, $w_{\alpha\beta}$ is just the number of the nearest neighbors
that belong to the set $M_\beta$ for $m_i\in M_\alpha$.
The critical anisotropy, where the $2\times 2$ determinant vanishes, is given
by
\begin{eqnarray}
\l_c=\f{A_1A_2}{A_1+2A_2}\approx 0.7157\,
\end{eqnarray}
which agrees well with the results of Wysin~\cite{Wysin1994}.
Wysin also showed the $\l_c$ is determined by the zero of a $3\times 3$
determinant as $0.7044$.
We obtain a smaller critical value of $\l_c\approx 0.6974$ for $\cN=4$, which
converges to $0.6941$ in four-digit accuracy for $\cN\ge 16$.
This critical value show a deviation from the result in numerical simulations.
Wysin showed that there are some differences in the discrete solution from the
continuum results especially near the vortex core~\cite{Wysin1998}.
Similar to Wysin's process, we eliminate this difference between discrete and
continuum solutions by evolution of each spin's $xy$ components to point along
the direction of the effective field due to its neighbors, and we obtain
$\l_c\approx0.703420$, which is very close to the value obtained by Wysin's,
$\l_c\approx 0.703409$, and one from our simulations in
Sec.~\ref{sec:simulations} ($\l_c\approx 0.7035$).

\begin{figure}[t] 
\begin{minipage}[b]{0.4cm} $\l_c$ \vspace{2.5cm} \end{minipage}
\begin{minipage}[b]{7cm}
  \includegraphics[width=7cm]{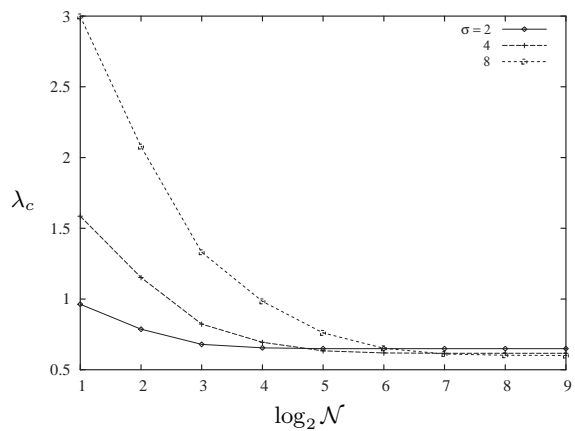} $\log_2 \cN$
\end{minipage}
\caption{ \label{fig:N_c}
  The critical anisotropy $\l_c$ as a function of $\log_2\cN$ for the
  anisotropic Heisenberg model with Gaussian type interactions for $\s=2,4$ and
  $8$.
  As $\s$ increases, the size of the out-of-plane vortex core increases.
}
\end{figure}

For the nearest-neighbor interactions in the CTDAHM, the size of the vortex
core is small, which leads to precise determination of the critical anisotropy
for relatively small $\cN$.
But as the interaction range increases, the size of the vortex core increases
and more components of the matrix are needed to obtain the accurate value of
$\l_c$.
Fig.~\ref{fig:N_c} shows the computed anisotropy $\l_c$ as a function of $\cN$
for different interaction ranges $\s_\mu$ in Gaussian type interactions.
The convergence of different lines for different ranges gives an estimate of
the vortex core size $r_v$.
For the average number of the degeneracy $g$ per block $M_\alpha$, $g\cN_c$
means the number of spin sites which contributes to the determination of $\l_c$.
The radius of the vortex core can be estimated as
\begin{eqnarray}
  r_v\sim \sqrt{\f{g}{\pi}\cN_c}
\end{eqnarray}
with $g$ approaching $8$ for large $\cN_c$ for the square lattice or the $D_4$
symmetry geometry.

%
%
\section{Single-Vortex Simulations}  \label{sec:simulations}
\setcounter{equation}{0}

\begin{figure}
\begin{minipage}[b]{6.0cm} \begin{center}
  \includegraphics[width=6.0cm]{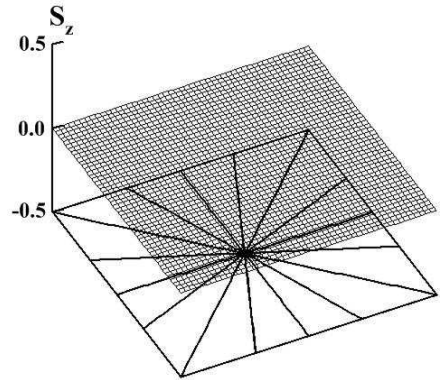} \\ (a)
\end{center} \end{minipage}
\\ 
\begin{minipage}[b]{6.0cm} \begin{center}
  \includegraphics[width=6.0cm]{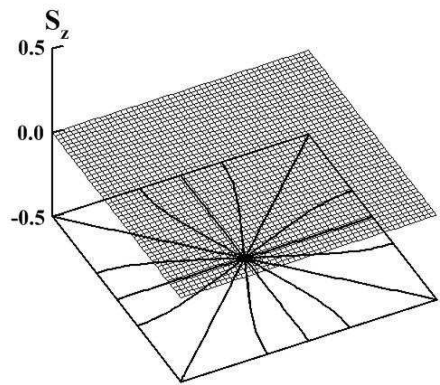} \\ (b)
\end{center} \end{minipage}
\\ 
\begin{minipage}[b]{6.0cm} \begin{center}
  \includegraphics[width=6.0cm]{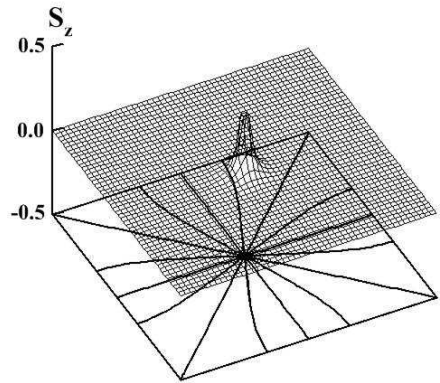} \\ (c)
\end{center} \end{minipage}
\caption{ \label{fig:single-vortex}
  Spin components near in-plane and out-of-plane vortex formed in the
  easy-plane Heisenberg model with Gaussian-type interactions.
  (a) The Initial state with an in-plane vortex at the lattice center with
  small random fluctuations of $|S_z|<10^{-6}$ and the states after evolution
  for (b) $\l=0.72$ and (c) $\l=0.74$ ($\l=\v_z/\v_{xy}$, $\s_{xy}=\s_z=1$,
  square lattice type and $50\times 50$ lattice size).
  In-plane spin components are represented by contour lines where
  $\phi_i=n\pi/8$ for $n=0,1\ldots 15$.
}
\end{figure}

\begin{figure}
\rotatebox{90} {
  \begin{minipage}[b]{6.0cm} $\sqrt{\langle (S_z)^2\rangle}$ \end{minipage}
}
\begin{minipage}[b]{7cm}
$\times 10^8$ \ \ \ \ \ \ \ \ \ \ \ \ \ \ \ \ \ \ \ \ \ \ \ \ \ \ \ \ \ \ \ \
\ \ \ \ \ \ \ \ \ \ \ \ \ \ \ \ \ \ \ \ \\
\includegraphics[width=7cm]{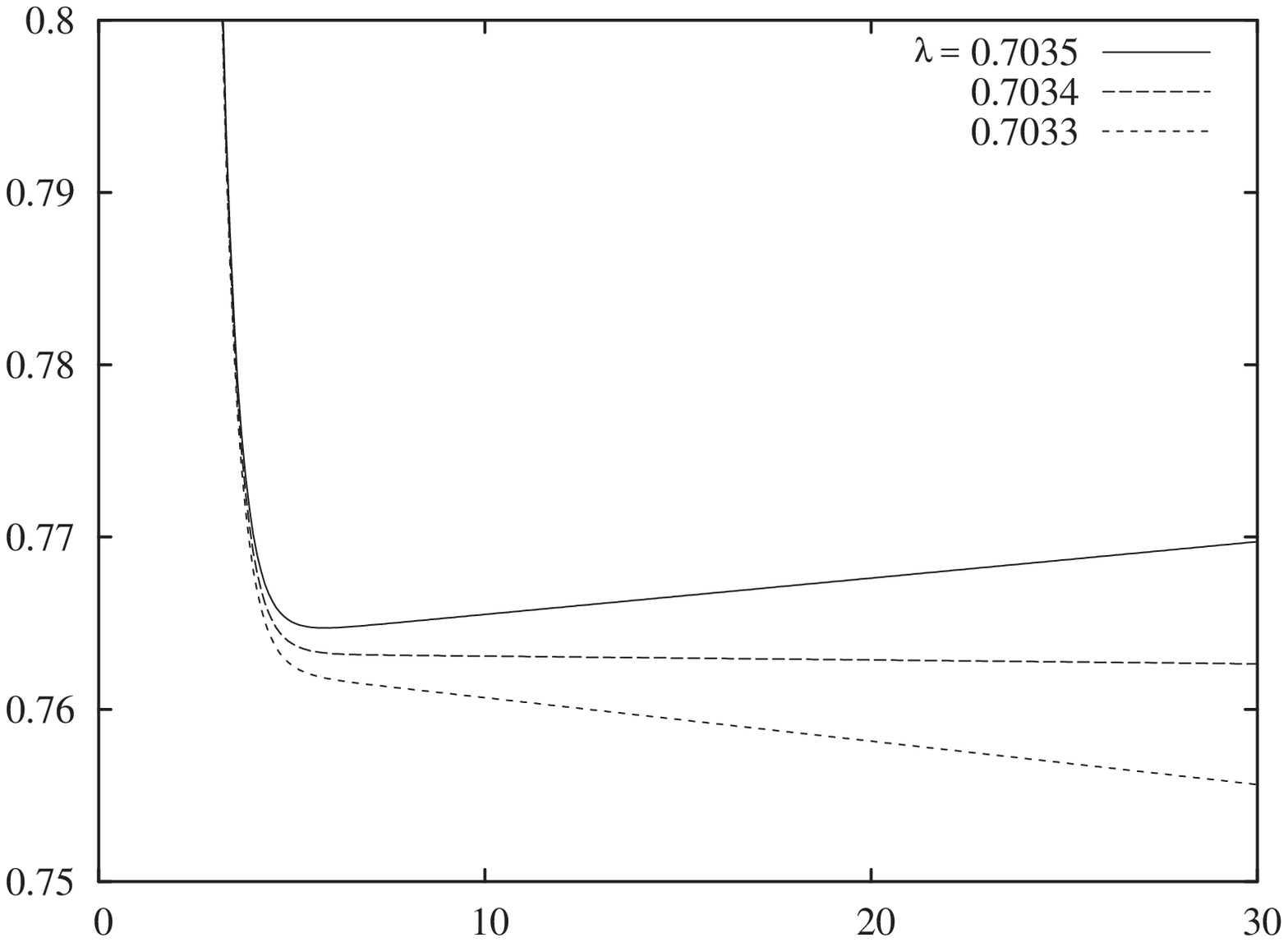} \\ \hspace{5.0cm} $t$ (sec)
\end{minipage}
\caption{ \label{fig:simulations}
  The growth or decay of the $S_z$ component in time for the CTDAHM in the case
  of single vortex simulation for the square lattice type ($50 \times 50$
  lattice size and time step of $0.001$).
}
\end{figure}

In general, it is quite difficult to determine the critical anisotropy
precisely with a simulated annealing approach using Monte Carlo simulations.
In particular, for the distance-dependent interaction models, the longer
annealing time is required due to the larger lattice size and the larger cutoff
range.
Our numerical simulations are carried out by evolving dynamical equations as a
gradient flow, which allows a more accurate determination of the vortex
behavior near $\l\sim\l_c$.
The discrete equations of motion used in simulations are 
\begin{subequations}
\begin{eqnarray} \label{eq:dynamic}
  \f{\partial\phi_i}{\partial t}&=&-\f{\partial H}{\partial\phi_i} \nonumber \\
  &=&-2\sum_j J_{xy}(r_{ij})\sin\theta_i\sin\theta_j\sin(\phi_i-\phi_j) \ \ \ \ \\
  \f{\partial\theta_i}{\partial t}&=&-\f{\partial H}{\partial\theta_i} \nonumber \\
  &=&2\sum_j\left\{J_{xy}(r_{ij})\cos\theta_i\sin\theta_j\cos(\phi_i-\phi_j)
    \right.  \nonumber \\
  && \hspace{1.5cm} \left.-J_z(r_{ij})\sin\theta_i\cos\theta_j \right\}
\end{eqnarray}
\end{subequations}
at zero temperature.
The simulations start from an initial state of a planar vortex on the lattice
center and small random fluctuations in out-of-plane components
($|S_z|<10^{-6}$).

A problem in the single vortex simulations is the effect of the area
boundaries.
The distribution of spin components is effected by the area boundary as shown
in Fig.~\ref{fig:single-vortex}.
At first, the contour lines of iso-phase are stretch out radially from the
vortex core.
But after dynamical evolution, they tend to meet perpendicularly with the area
boundary due to the equilibrium condition $\d H/\d\phi\sim 0$ or
$\nabla^2\phi\sim 0$.
If the lattice size is not sufficiently large, the distribution of spin
components near the vortex core is effected by the area boundary.
For the spin system with distance-dependent interactions, the size of the
vortex core increases, so that the larger lattice size is necessary for
simulations.
Note also that the shape of the lattice is matter more than its size.
Note that the single vortex is attracted to the area boundary like an electric
charge near the conductor surface.
If there is an asymmetry in the lattice, the centered vortex may move along a
certain direction.
An asymmetry in the shape of the cutoff range can also induce the wandering of
the vortex core. 
Especially the single-vortex simulations for the triangular lattice turned out
to be more difficult and sensitive to the shape of the lattice and the cutoff
range.

We measure the growth of the out-of-plane components by
$\sqrt{\langle(S_z)^2\rangle}$.
The measurement of the squared component $(S_z)^2$ at the central peak can be
another choice. 
However, tracing single or few components may fail to determine the vortex
stability since the vortex core may start to move. 
In case that the vortex wanders, the fluctuations can be measured by
$\sqrt{\langle (S_z)^2\rangle}$.
To clarify the simulation results, we confirm the monotonical increase or
decrease after the inflection point.

Our simulational method is very fast, yielding the computed critical anisotropy
that is more accurate than those by previous approaches.
For example, Fig.~\ref{fig:simulations} shows the results of dynamical
evolution with the CTDAHM for the square lattice.
We find a monotonical growth (or decay) of out-of-plane components after some
time for $\l=0.7035$ (or $\l=0.7034$, $0.7033$).
This graph suggests that the critical anisotropy lies between
$0.7034 < \l_c \le 0.7035$, which agrees well with analytic calculations by
Wysin in Ref.~9 ($\l_c\approx 0.703409$) and ours in Sec.\ref{sec:analytic}
($\l_c\approx 0.703420$).
Similarly, we obtain $\l_c\approx 0.6129$ for the triangular lattice and
$\l_c\approx 0.8330$ for the hexagonal lattice, which also agree well with the
results of Wysin ($\l_c\approx 0.612856$ and $0.832956$
respectively~\cite{Wysin1998}).

%
%
\section{Experiments in Gaussian type interactions} \label{sec:gaussian}
\setcounter{equation}{0}

\begin{figure}
\begin{minipage}[b]{7.0cm} \begin{center}
  \includegraphics[width=7.0cm]{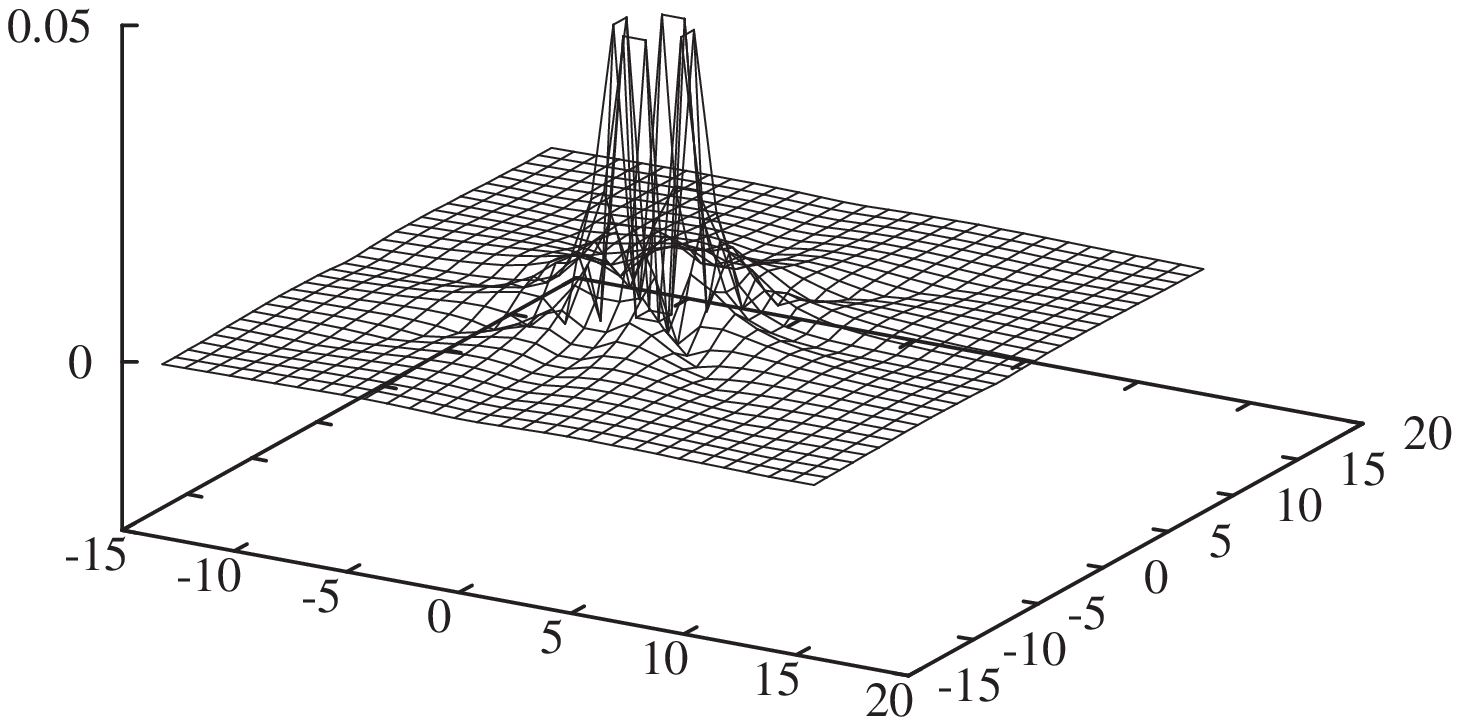} \\ (a)
\end{center} \end{minipage}
\\ \ \\
\begin{minipage}[b]{7.0cm} \begin{center}
  \includegraphics[width=7.0cm]{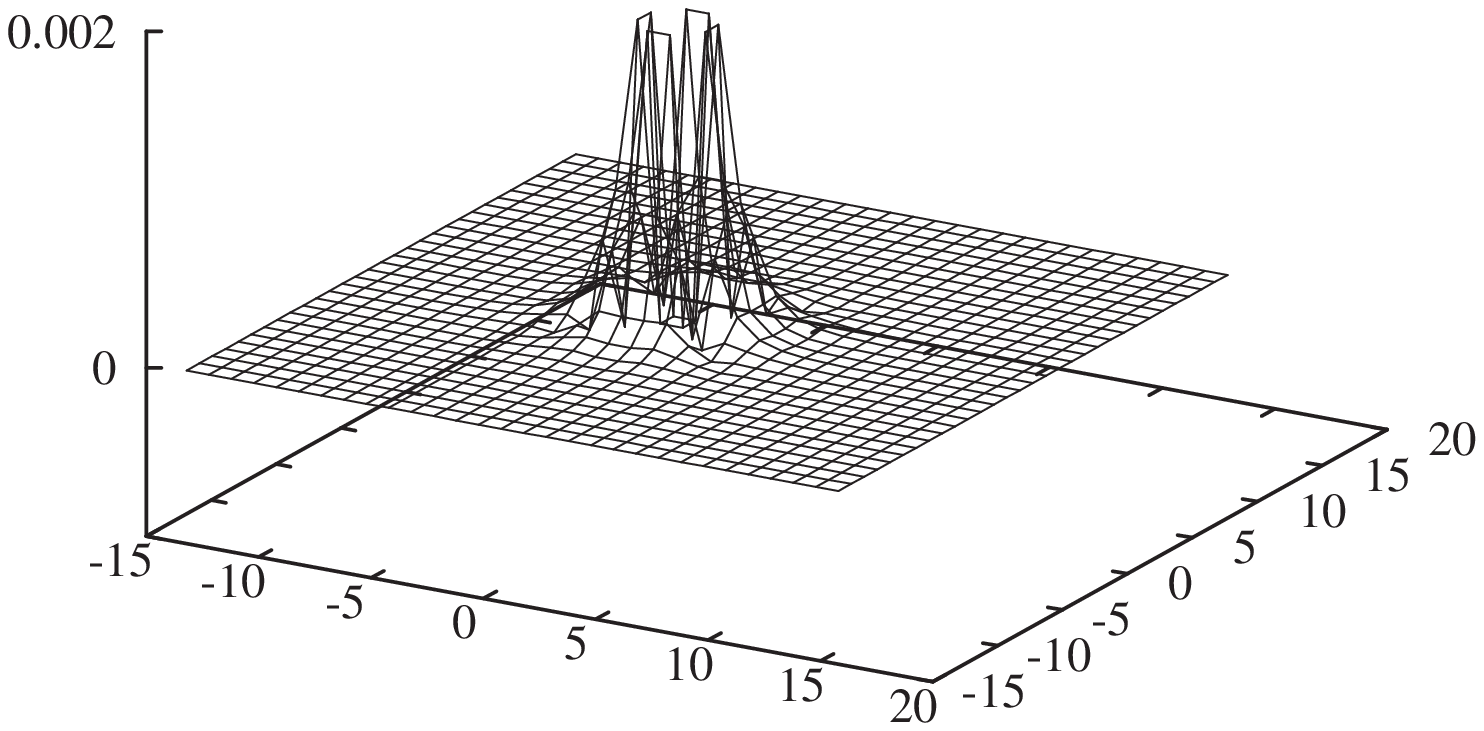}
  \\ (b)
\end{center} \end{minipage}
\\ \ \\
\begin{minipage}[b]{7.0cm} \begin{center}
  \includegraphics[width=7.0cm]{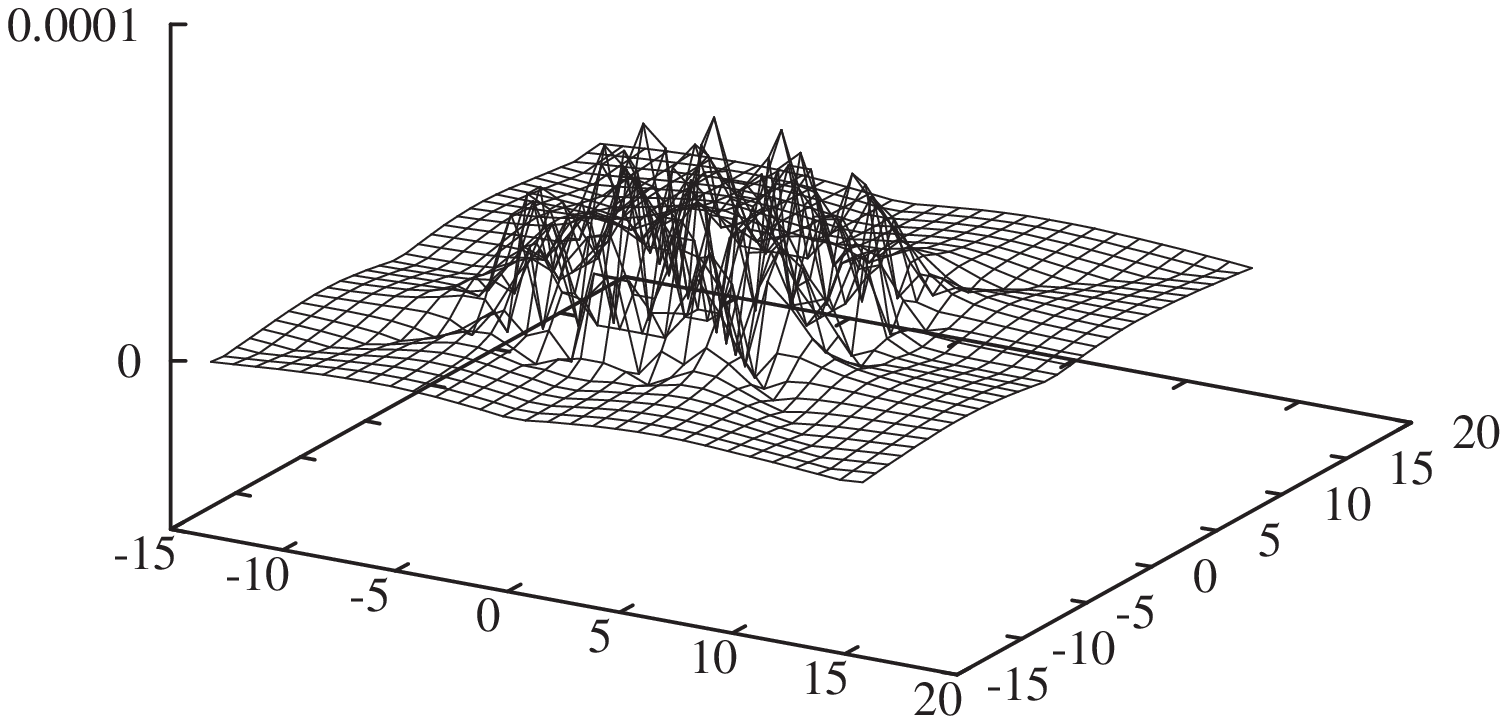}
  \\ (c)
\end{center} \end{minipage}
\caption{ \label{fig:diff}
  Deviation in the discrete solution from the continuum result near the vortex
  core for the square lattice, where the $x$- and $y$-axis are the lattice
  coordinates and the z-axis denotes $|\sin(\phi_{dis}-\phi_{cont})|$ for
  (a) the nearest-neighbor interactions (b) the Gaussian type interactions with
  $\s_{xy}=1$ and (c) the Gaussian type interactions with $\s_{xy}=4$.
}
\end{figure}

As a typical example of the spin systems with an arbitrary distance-dependent
interaction types, we study the model with Gaussian type interactions where
\begin{eqnarray} \label{eq:gaussian}
  I_\mu(r)=\exp(-r^2/2\s_\mu^2).
\end{eqnarray}
In the continuum limit, the fourier transformed interaction,
\begin{eqnarray}
  \tI_\mu(q)=\pi\left(\f{\s_\mu}{a}\right)^2\exp\left(-\s_\mu^2q^2/2\right),
\end{eqnarray}
has a maximum at $q=0$ and the approximated Hamiltonian can be written as
Eq.(\ref{eq:approx}) with $J_s=\pi\v_{xy}\s_{xy}^2/a^2$,
$\d=1-(\v_z/\v_{xy})(\s_z^2/\s_{xy}^2)$,
$\d'=1-(\v_z/\v_{xy})(\s_z^4/\s_{xy}^4)$, $\alpha^2=2\s_{xy}^2$ and the
anisotropy
\begin{eqnarray}
  \l=\f{\v_z}{\v_{xy}}\f{\s_z^2}{\s_{xy}^2}.
\end{eqnarray}
The critical anisotropy $\l_c$ is not uniform but varies depending on $\v_\mu$
and $\s_\mu$.
We can investigate the behaviour of the critical anisotropy from Gaussian type
interactions for two different general cases; the exchange strengths are
proportional to each other, $J_z(r)/J_{xy}(r)=const$ for all $r$ or
$\l=\v_z/\v_{xy}$, and the exchange energy are proportional to $\s_\mu^2$ for
the interaction range $\s_\mu$ in the two-dimensional system or
$\l=\s_z^2/\s_{xy}^2$.

\begin{figure}
\begin{minipage}[b]{0.5cm} (a) \vspace{3.5cm} \end{minipage}
\begin{minipage}[b]{0.4cm} $\l_c$ \vspace{2.5cm} \end{minipage}
\begin{minipage}[b]{7cm}
  \includegraphics[width=7cm]{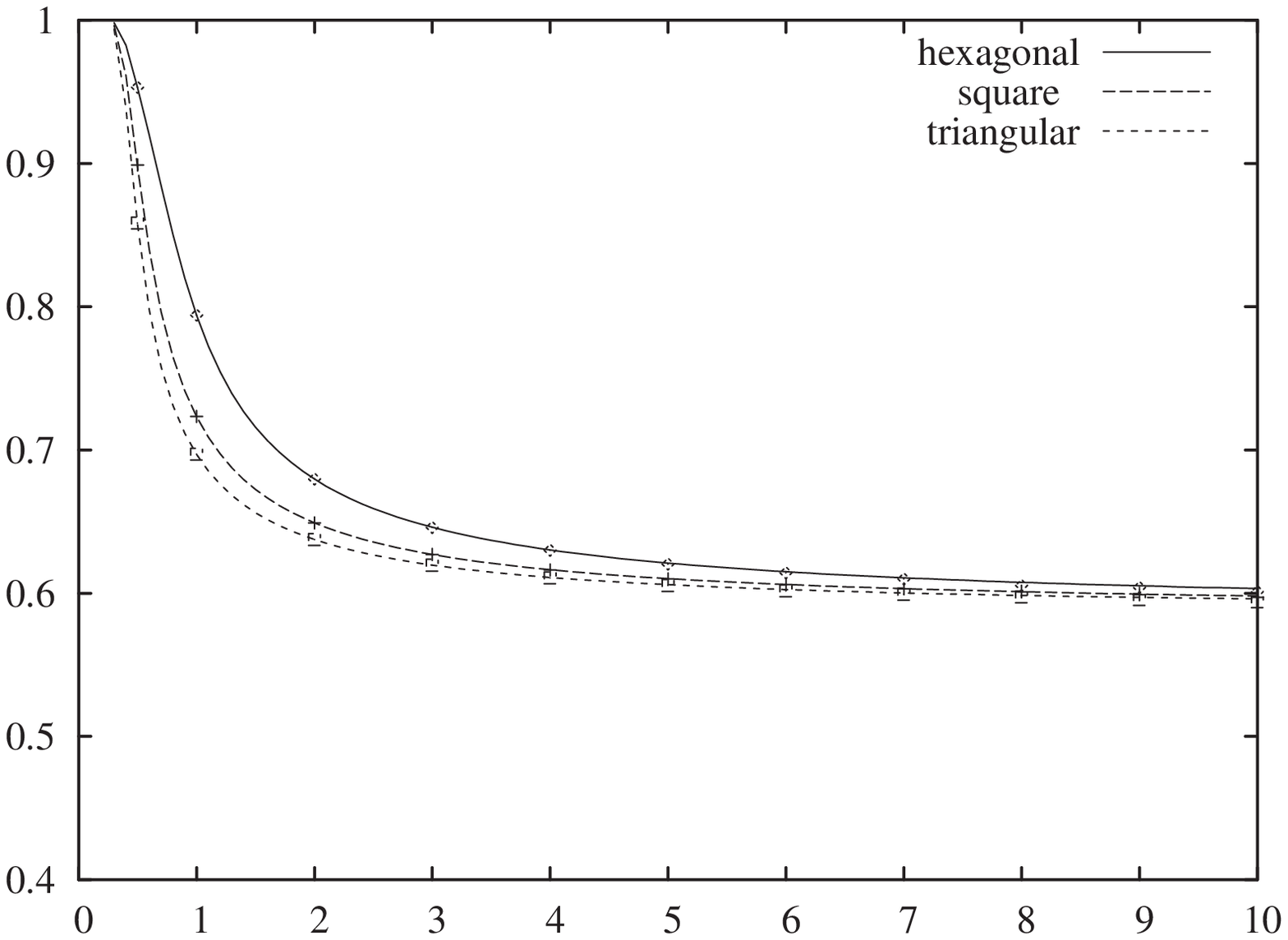} $\s_{xy}$
\end{minipage}
\\
\begin{minipage}[b]{0.5cm} (b) \vspace{3.5cm} \end{minipage}
\begin{minipage}[b]{0.4cm} $\l_c$ \vspace{2.5cm} \end{minipage}
\begin{minipage}[b]{7cm}
  \includegraphics[width=7cm]{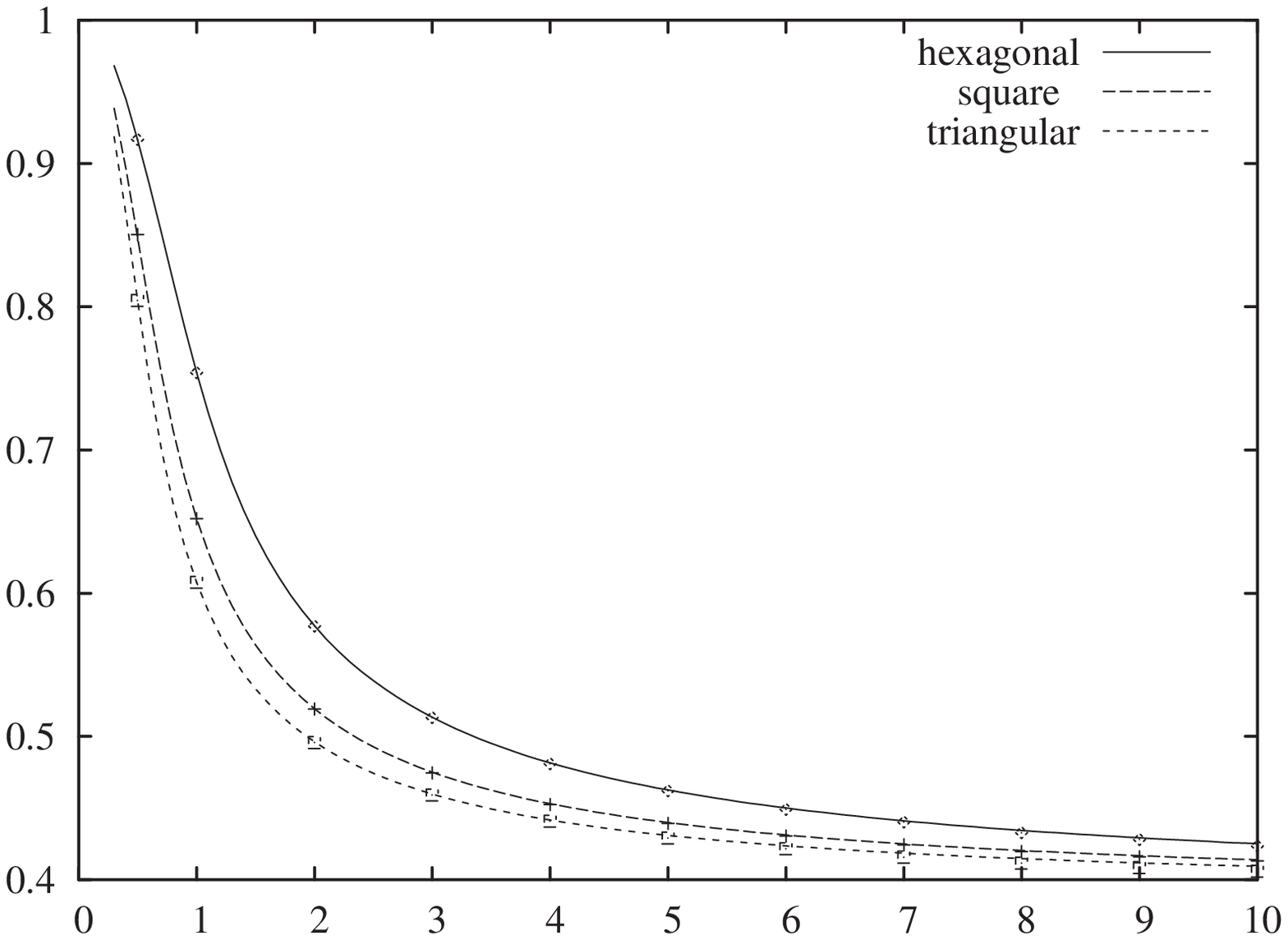} \\ $\s_{xy}$
\end{minipage}
\caption{ \label{fig:l_c}
  The critical anisotropy $\l_c$ as a function of $\s_{xy}$ in the anisotropic
  Heisenberg model with Gaussian type interactions for (a) $\l=\v_z/\v_{xy}$
  ($\s_{xy}=\s_z$) and (b) $\l=\s_z^2/\s_{xy}^2$ ($\v_{xy}=\v_z$).
  The critical anisotropy $\l_c$ are determined by the determinant of the
  $1024\times 1024$ matrix.
  The symbols denote the results of the numerical simulations.
}
\end{figure}

In analytic calculations, a large number of the nearest components, $\cN$, is
needed for the convergence of the critical value $\l_c$ as shown in
Fig.~\ref{fig:N_c}.
The critical values of the anisotropy are determined for $\cN=1024$.
For the case of $\l=\v_z/\v_{xy}$, the critical anisotropy is found by the
condition that the determinant of the $\cN\times\cN$ matrix, $W$, vanishes
where
\begin{eqnarray} \label{eq:W_v}
  W_{\alpha\beta}&=&\d_{\alpha\beta}\sum_{i\in M_\alpha}\sum_j
    \exp(-r_{ij}^2/2\s_{xy}^2)\cos(\phi_i-\phi_j) \nonumber \\
    &&-\l\sum_{i\in M_\alpha}\sum_{j\in M_\beta}\exp(-r_{ij}^2/2\s_{xy}^2).
\end{eqnarray}
The maximal eigenvalue $\l$ of Eq.(\ref{eq:W_v}) gives the critical anisotropy
$\l_c$ as in Eq.(\ref{eq:eigenproblem}).
The angles of in-plane components $\phi_i$ are initially set to their continuum
vortex values, $\phi_i=\tan^{-1}(y_i/x_i)$, for each site $i$ and evolved until
they approach the equilibrium state according to the dynamical equation,
\begin{eqnarray} \label{eq:op_evol}
  \f{\partial\phi_i}{\partial t} &=&-2\sum_j J_{xy}(r_{ij})\sin(\phi_i-\phi_j)
\end{eqnarray}
from Eq.(\ref{eq:dynamic}) with $m_i=0$ for all sites, that is, the discrete
dynamics in the absence of out-of-components or the pure XY model limit.
Note that the spin configuration of the system with distance interaction,
however, corresponds well to the continuum results near the vortex core in
contrast to one with the nearest-neighbor interaction.
There are much differences between our results and the classical models with
the nearest-neighbor interactions, which decreases as the interaction range
increases for the Gaussian type interactions.
Fig.~\ref{fig:diff} shows the differences between the discrete solutions and
the continuum solutions.
The discrete solutions are obtained after the evolution from the continuum
solution $\phi_i=\tan^{-1}(y_i/x_i)$ by Eq.(\ref{eq:op_evol}), avoiding the
area boundary effect by using a round shaped lattice boundary ($R=80$).
However, we solve the equations with the initial angles or setting
$\cos(\phi_i-\phi_j)=\hat{r}_i\cdot\hat{r}_j$ for Gaussian type interactions,
and which leads to the accurate results.
For $\l=\s_z^2/\s_{xy}^2$, the matrix $W$ is given by
\begin{eqnarray} \label{eq:W_s}
  W_{\alpha\beta}&=&\d_{\alpha\beta}\sum_{i\in M_\alpha}\sum_j
    \exp(-r_{ij}^2/2\s_{xy}^2)\cos(\phi_i-\phi_j) \nonumber \\
    &&-\sum_{i\in M_\alpha}\sum_{j\in M_\beta}\exp(-r_{ij}^2/2\l\s_{xy}^2),
\end{eqnarray}
which is difficult to be converted as an eigenvalue problem as before.
The critical anisotropy $\l_c$ is determined by $f(\l)=\det|W(\l)|$ in
Eq.(\ref{eq:W_s}) using the root finding method, which is in agreement with the
results of numerical simulations within the accuracy of two- or three-digits.
We show the analytic and simulational results for square, hexagonal and
triangular lattices in Fig.~\ref{fig:l_c}.

\begin{figure}
\rotatebox{90} { \begin{minipage}[b]{8.0cm}
  $\sigma_{xy}\sqrt{\lambda_c}$
\end{minipage} }
\begin{minipage}[b]{7cm}
  \includegraphics[width=7cm]{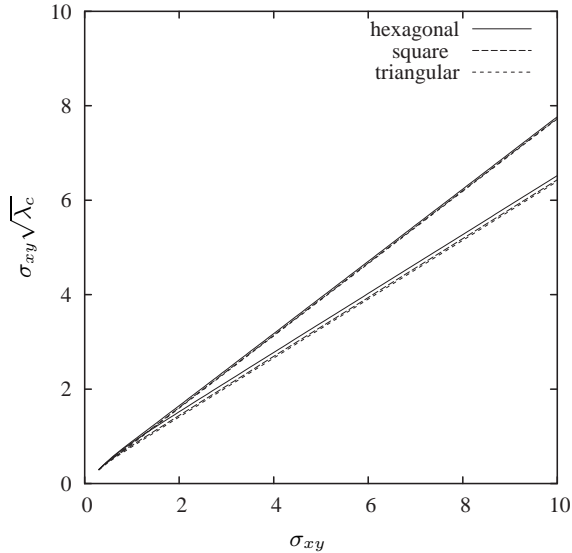} $\sigma_{xy}$
\end{minipage}
\caption{ \label{fig:linear}
  $\s_{xy}\sqrt{\l_c}$ as a function of $\s_{xy}$.
  The upper branch corresponds to $\l=\v_z/\v_{xy}$ and lower branch to
  $\l=\s_z^2/\s_{xy}^2$.
  When $\l=\s_z^2/\s_{xy}^2$, y-axis means $\s_z$.
}
\end{figure}

\begin{table}
\caption{ \label{table:linear}
  Linear fitting of $\s_{xy}\sqrt{\l_c}$ as a function of $\s_{xy}$
}
\begin{tabular}{|c|cc|cc|}
\hline
  Lattice
  & \multicolumn{2}{c|}{$\l=\v_z/\v_{xy}$}
  & \multicolumn{2}{c|}{$\l=\s_z^2/\s_{xy}^2$} \\ \cline{2-5}
  type & \ slope \ & y-intercept & \ slope \ & y-intercept \\ \hline
Square         & 0.764928 & 0.082300 & 0.625910 & 0.181816 \\
 Hexagonal     & 0.765017 & 0.116011 & 0.629297 & 0.243578 \\
\ Triangular \ & 0.764967 & 0.068548 & 0.624828 & 0.154744 \\
\hline
\end{tabular}
\end{table}

The obtained critical value of anisotropy decreases as the interactions range
$\s$ decreases.
Note that the computed values of the critical anisotropy $\l_c$ are in the
order of hexagonal, square and triangular lattices type for both
$\l=\v_z/\v_{xy}$ and $\l=\s_z^2/\s_{sy}^2$ as in the CTDAHM.
Fig.~\ref{fig:linear} shows the plot of $\s_{xy}\sqrt{\l_c}$ as a function of
$\s_{xy}$, showing a linearity between them.
The slopes for different lattice types can be grouped as two classes with
$\l=\v_z/\v_{xy}$ and $\l=\s_z^2/\s_{xy}^2$ (Table~\ref{table:linear}).
If the linearity is maintained for all $\s_{xy}$, the asymptotic value of
$\sqrt{\l_c}$ will be converge to the slope of the line for large $\s_{xy}$.
With the help of the asymptotic solutions in Eq.(\ref{eq:W_v}) and
Eq.(\ref{eq:W_s}), we can guess that the critical anisotropy $\l_c$ will
converges to zero and the slope of lines in Fig.~\ref{fig:linear} will become
smaller for large $\s_{xy}$

When $\v_{xy}=\v_z$ and $\s_{xy}=\s_z=1/\sqrt{2}$, the approximated Hamiltonian with Gaussian type
interactions in Eq.(\ref{eq:approx}) becomes that
\begin{eqnarray}
  H&\simeq&-NJ_sS^2+\f{J_sS^2}{4}\left.\int d^2r\right\{(1-m^2)(\nabla\phi)^2
    \nonumber \\
  && \ \ \left.+[1-\d(1-m^2)]\f{(\nabla m)^2}{(1-m^2)}+4\d m^2 \right\},
\end{eqnarray}
which is equivalent to that of the CTDAHM~\cite{Gouvea1989} with a coupling
constant of nearest-neighbor interactions $J_s/2$.
In this case, they have the common ground-state energy and the common excitation
energy due to formation of a vortex for the same lattice type and lattice size.
However, we find that the critical anisotropy $\l_c$ in Gaussian type
interactions is 0.7938 from analytic calculations and $0.7941$ from
simulational experiments for square lattice, which are different with
$\l_c\approx 0.7035$ in the CTDAHM.
These results mean that the solutions of the critical anisotropy from the
continuum limit approach are inadequate.

%
%
\section{Discussion}
\setcounter{equation}{0}

We have studied the vortex formation in the two-dimensional anisotropic
Heisenberg model with distance-dependent interactions and determined the
critical anisotropy $\l_c$ precisely for Gaussian type interactions.
The continuum theory helps to predict how spin configurations will develop
for an arbitrary shaped exchange function $J_\mu(r)$.
If the exchange energy function $J_\mu(r)$ satisfies certain conditions, the
approximated Hamiltonian in the continuum limit hows a similarity to that of
the CTDAHM, which leads to the similarity in two types of static vortex
solutions.
However, the continuum theory breaks down in determining the critical
anisotropy as noted by Wysin~\cite{Wysin1994}.
We also give an example that two different anisotropic Heisenberg models with a
common approximated form in the continuum limit can have different critical
anisotropies.

\begin{figure}[t]
\begin{minipage}[b]{0.5cm} $\s_z$ \vspace{5cm} \end{minipage}
\begin{minipage}[b]{6cm}
  \includegraphics[width=6cm]{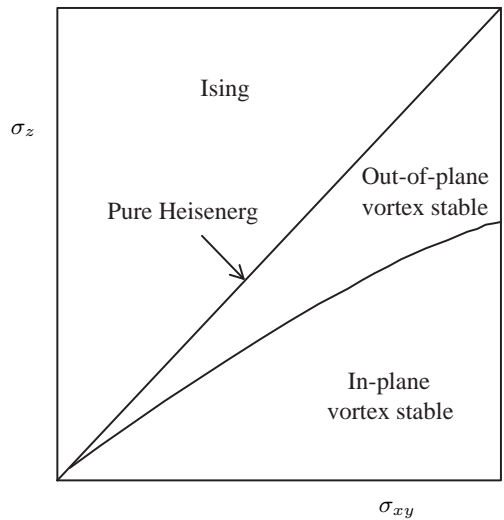} \\ \hspace{3cm} $\s_{xy}$
\end{minipage}
\caption{ \label{fig:phase}
  Crossover behavior for interaction ranges.
}
\end{figure}

It is very difficult to determine the critical anisotropy for an arbitrary type
of the exchange energy function $J_\mu(r)$.
In this paper, we have studied two typical cases by using Gaussian type
interaction models.
One case is that $J_{xy}(r)$ and $J_z(r)$ are proportional for all $r$ and
another case is that $J_{xy}(r)$ and $J_z(r)$ have a common distribution
type with different interaction ranges.
The results for these cases suggest some general behaviors of $\l_c$. 
The critical anisotropy is the smallest for the triangular lattice and is the
largest for the hexagonal lattice among three lattice types explored.
This is also the case for the classical model because the spins near the vortex
core tend to be more parallel, reducing their total exchange energy further, so
that the lattice type with more nearest neighborhoods can reduce the
out-of-plane exchange energy further while increasing the in-plane exchange
energy for lower value of $\l$.
Fig.~\ref{fig:phase} shows the crossover behavior of spin dynamics when the
anisotropy between the in-plane and the out-of-plane energies is manifested
only by the difference in interaction ranges.
If the interaction range of the in-plane is larger than that of the
out-of-plane ($\s_{xy}>\s_z$), the number of exchange pairs in the in-plane
case is larger for the ferromagnetic system and the system belong to the same
universality class as the XY model.
The region where the out-of-plane vortex is stable can be approximately found
in the region with $\alpha<\s_z/\s_{xy}<1$, for a reasonable range of $\s$.
If $\s_{xy}$ is very large in comparison with the nearest-neighbor distance,
a nearly zero critical anisotropy may occur.

Our study for the precise determination of the critical anisotropy is motivated
from the problem of the map formation in the cerebric cortex rather than
magnetism systems.
The results for the crossover among the regions with Ising, stable out-of-plane
or stable in-plane vortex behaviors depending on the interaction ranges in
Fig.~\ref{fig:phase} correspond well with the data from the animal experiments
for different types of the map formation in the visual cortex~\cite{Cho2003B}.

\bibliography{prb}

\begin{thebibliography}{11}
\expandafter\ifx\csname natexlab\endcsname\relax\def\natexlab#1{#1}\fi
\expandafter\ifx\csname bibnamefont\endcsname\relax
  \def\bibnamefont#1{#1}\fi
\expandafter\ifx\csname bibfnamefont\endcsname\relax
  \def\bibfnamefont#1{#1}\fi
\expandafter\ifx\csname citenamefont\endcsname\relax
  \def\citenamefont#1{#1}\fi
\expandafter\ifx\csname url\endcsname\relax
  \def\url#1{\texttt{#1}}\fi
\expandafter\ifx\csname urlprefix\endcsname\relax\def\urlprefix{URL }\fi
\providecommand{\bibinfo}[2]{#2}
\providecommand{\eprint}[2][]{\url{#2}}

\bibitem[{\citenamefont{Berezinskii}(1971)}]{Berezinskii1971}
\bibinfo{author}{\bibfnamefont{V.~L.} \bibnamefont{Berezinskii}},
  \bibinfo{journal}{Teor.\ Fiz.} \textbf{\bibinfo{volume}{61}},
  \bibinfo{pages}{1144} (\bibinfo{year}{1971}).

\bibitem[{\citenamefont{Kosterlitz and Thouless}(1973)}]{Kosterlitz1973}
\bibinfo{author}{\bibfnamefont{J.~M.} \bibnamefont{Kosterlitz}}
  \bibnamefont{and} \bibinfo{author}{\bibfnamefont{D.~J.}
  \bibnamefont{Thouless}}, \bibinfo{journal}{J.\ Phys.\ C}
  \textbf{\bibinfo{volume}{6}}, \bibinfo{pages}{1181} (\bibinfo{year}{1973}).

\bibitem[{\citenamefont{Kosterlitz}(1974)}]{Kosterlitz1974}
\bibinfo{author}{\bibfnamefont{J.~M.} \bibnamefont{Kosterlitz}},
  \bibinfo{journal}{J.\ Phys.\ C} \textbf{\bibinfo{volume}{7}},
  \bibinfo{pages}{1046} (\bibinfo{year}{1974}).

\bibitem[{\citenamefont{Wysin et~al.}(1988)\citenamefont{Wysin, Gouv{\^e}a,
  Bishop, and Mertens}}]{Wysin1988}
\bibinfo{author}{\bibfnamefont{G.~M.} \bibnamefont{Wysin}},
  \bibinfo{author}{\bibfnamefont{M.~E.} \bibnamefont{Gouv{\^e}a}},
  \bibinfo{author}{\bibfnamefont{A.~R.} \bibnamefont{Bishop}},
  \bibnamefont{and} \bibinfo{author}{\bibfnamefont{F.~G.}
  \bibnamefont{Mertens}}, in \emph{\bibinfo{booktitle}{Computer Simulations
  Studies in Condensed Matter Physics}}, edited by
  \bibinfo{editor}{\bibfnamefont{D.~P.} \bibnamefont{Landau}},
  \bibinfo{editor}{\bibfnamefont{K.~K.} \bibnamefont{Mon}}, \bibnamefont{and}
  \bibinfo{editor}{\bibfnamefont{H.-B.} \bibnamefont{Sch{\"u}tter}}
  (\bibinfo{publisher}{Spinger-Verlag, Berln}, \bibinfo{year}{1988}).

\bibitem[{\citenamefont{Gouv{\^e}a et~al.}(1989)\citenamefont{Gouv{\^e}a,
  Wysin, and Bishop}}]{Gouvea1989}
\bibinfo{author}{\bibfnamefont{M.~E.} \bibnamefont{Gouv{\^e}a}},
  \bibinfo{author}{\bibfnamefont{G.~M.} \bibnamefont{Wysin}}, \bibnamefont{and}
  \bibinfo{author}{\bibfnamefont{A.~R.} \bibnamefont{Bishop}},
  \bibinfo{journal}{Phys.\ Rev.\ B} \textbf{\bibinfo{volume}{39}},
  \bibinfo{pages}{11840} (\bibinfo{year}{1989}).

\bibitem[{\citenamefont{Costa and Costa}(1996)}]{Costa1996}
\bibinfo{author}{\bibfnamefont{J.~E.~R.} \bibnamefont{Costa}} \bibnamefont{and}
  \bibinfo{author}{\bibfnamefont{B.~V.} \bibnamefont{Costa}},
  \bibinfo{journal}{Phys.\ Rev.\ B} \textbf{\bibinfo{volume}{54}},
  \bibinfo{pages}{994} (\bibinfo{year}{1996}).

\bibitem[{\citenamefont{Wysin}(1994)}]{Wysin1994}
\bibinfo{author}{\bibfnamefont{G.~M.} \bibnamefont{Wysin}},
  \bibinfo{journal}{Phys.\ Rev.\ B} \textbf{\bibinfo{volume}{49}},
  \bibinfo{pages}{8780} (\bibinfo{year}{1994}).

\bibitem[{\citenamefont{Cho and Kim}(2003{\natexlab{a}})}]{Cho2003A}
\bibinfo{author}{\bibfnamefont{M.~W.} \bibnamefont{Cho}} \bibnamefont{and}
  \bibinfo{author}{\bibfnamefont{S.}~\bibnamefont{Kim}},
  \bibinfo{journal}{Phys.\ Rev.\ Lett. (to be published)}
  (\bibinfo{year}{2003}{\natexlab{a}}), \bibinfo{note}{arXiv:Physics/0306047}.

\bibitem[{\citenamefont{Cho and Kim}(2003{\natexlab{b}})}]{Cho2003B}
\bibinfo{author}{\bibfnamefont{M.~W.} \bibnamefont{Cho}} \bibnamefont{and}
  \bibinfo{author}{\bibfnamefont{S.}~\bibnamefont{Kim}},
  \emph{\bibinfo{title}{Different ocular dominance map formation influenced by
  orientation preference columns}} (\bibinfo{year}{2003}{\natexlab{b}}),
  \bibinfo{note}{arXiv:q-bio.NC/0310039}.

\bibitem[{\citenamefont{Takeno and Homma}(1980)}]{Takeno1980}
\bibinfo{author}{\bibfnamefont{S.}~\bibnamefont{Takeno}} \bibnamefont{and}
  \bibinfo{author}{\bibfnamefont{S.}~\bibnamefont{Homma}},
  \bibinfo{journal}{Theor.\ Phys.} \textbf{\bibinfo{volume}{64}},
  \bibinfo{pages}{1193} (\bibinfo{year}{1980}).

\bibitem[{\citenamefont{Wysin}(1998)}]{Wysin1998}
\bibinfo{author}{\bibfnamefont{G.~M.} \bibnamefont{Wysin}},
  \bibinfo{journal}{Phys.\ Lett.\ A} \textbf{\bibinfo{volume}{23}},
  \bibinfo{pages}{96} (\bibinfo{year}{1998}).

\end{thebibliography}

\end{document}